\documentclass[12pt]{article}

\usepackage{a4wide}
\usepackage[pdftex,usenames,dvipsnames]{color}
\usepackage{graphics}
\usepackage{amsfonts}
\usepackage{amssymb}
\usepackage{accents}

\newcommand{\nit}{\noindent}

\newcommand{\np}{\newpage}
\newcommand{\dsp}{\displaystyle}
\newcommand{\vs}[1]{\vspace{#1 ex}}
\newcommand{\hs}[1]{\hspace{#1 em}}
\newcommand{\bfr}{\begin{flushright}}
\newcommand{\efr}{\end{flushright}}
\newcommand{\bc}{\begin{center}}
\newcommand{\ec}{\end{center}}
\newcommand{\ben}{\begin{enumerate}}
\newcommand{\een}{\end{enumerate}}

\newcommand{\be}{\begin{equation}}
\newcommand{\ee}{\end{equation}}
\newcommand{\ba}{\begin{array}}
\newcommand{\ea}{\end{array}}
\newcommand{\ct}{\cite}
\newcommand{\bit}{\bibitem}

\newcommand{\eps}{\epsilon}
\newcommand{\ve}{\varepsilon}

\newcommand{\kg}{\kappa}
\newcommand{\lb}{\lambda}
\newcommand{\sg}{\sigma}

\newcommand{\vf}{\varphi}
\newcommand{\og}{\omega}

\newcommand{\Del}{\Delta}

\newcommand{\hvf}{\hat{\vf}}

\newcommand{\lh}{\left(}
\newcommand{\rh}{\right)}
\newcommand{\ld}{\left.}

\newcommand{\cotanh}{\mbox{\,ctgh\,}}

\begin{document}

\pagestyle{empty}

\vs{7}
\bc
{\Large \bf Dynamics of cosmological scalar fields$^*$}
\vs{5}

{\large J.W. van Holten}
\vs{3}

Nikhef, Amsterdam NL
\vs{2}

and
\vs{2} 

Leiden University, Leiden NL
\vs{4}

May 23, 2023
\ec
\vs{7}

{\small \nit
{\bf Abstract} \\
This paper reviews the dynamics of an isotropic and homogeneous cosmological scalar field. 
A general approach to the solution of the Einstein-Klein-Gordon equations is developed, which 
does not require slow-roll or other approximations. General conclusions about the qualitative 
behaviour of the solutions can be drawn, and examples of explicit solutions for some interesting 
cases are given. It is also shown how to find scalar potentials giving rise to a predetermined 
scalar field behaviour and associated evolution of the scale factor. }

\vfill 
\nit
{\footnotesize $^*$ This paper is dedicated to Richard Kerner on the occasion of his 80th anniversary.}

\np

\pagestyle{plain} 
\pagenumbering{arabic}

\nit
\section{Introduction and motivation \label{s.1}}

Observations of variable stars, such as type Ia supernovae, and the distribution of galaxies
in the universe, as well as the detailed maps of the Cosmic Microwave Background (CMB) 
made by the successive COBE, WMAP and PLANCK missions have provided impressive 
amounts of data which allow us to trace the evolution of the universe over the last 13.8 billion 
years, back to a time at which the universe was more than 1000 times smaller in all directions \ct{cobe,wmap,planck}. What those studies have taught us is, that for most of that past the rate 
of cosmic expansion was determined by the density of massive matter in the universe. About 
one seventh of that matter is of baryonic type, mostly hydrogen and helium. The rest is of a 
different, unknown but electrically neutral type; it is generally referred to as dark matter. 

Surprisingly however, since a bit more than four billion years ago the rate of cosmic 
expansion has become dominated by another source of energy density, which is not some 
kind of massive matter \ct{snls:2014,DES:2021}. In view of its unknown nature, and as its 
makes its presence known exclusively via the accelerated rate of expansion of the universe, 
it is referred to as dark energy. In the context of General Relativity (GR) it can be parametrized 
by a cosmological constant. But as there is no information on the evolution of dark energy from 
earlier times, it is impossible to say if it has always been really constant, or whether the dark 
energy density evolves itself over cosmological times. Another possible scenario would be 
for the cosmic expansion to be driven in part by one or more cosmic scalar fields, the energy 
of which can vary during the evolution of the universe. Such scenarios have been widely 
discussed, e.g.\ in the context of cosmic quintessence models \ct{wetterich:1988, zlatev:1999}. 

Another epoch of accelerating expansion of the universe is thought to have taken place 
in a very early phase to explain the present homogeneity and isotropy of the cosmos, 
and in particular of the CMB in which fluctuations are smaller than one part in $10^4$. 
This era of cosmic inflation must have smoothed out any inhomogeneities in the 
distribution of matter and radiation in the part of the universe within our cosmic horizon
\ct{starobinsky:1980}-\ct{turner:1983}. Its origin is likewise to be traced to some source 
of cosmic vacuum energy for which a scalar field, refered to as the inflaton field, may 
have been responsible.

As the discovery of the Higgs particle has made clear, scalar fields also play an important 
role in particle physics, where the Brout-Englert-Higgs (BEH) field is responsible for the 
masses of weak vector bosons and charged leptons, and contributes to those of quarks. 
The mechanism behind the BEH-effect is the non-zero vacuum expectation value (vev) 
acquired by the scalar field which couples to particle fields with no mass to begin with. 
The vev of the scalar field is however temperature dependent, and can therefore change 
over the life time of the  universe. This suggests there may have been a period in the 
very early universe at extremely high temperature during which there was not yet a vev 
of the BEH-field, and leptons and vector bosons were all massless. Such a scenario 
then creates the problem, that the energy density of the BEH-field at very early times 
would have been hugely different from the present (in which it contributes virtually 
nothing to the cosmic energy density). Presumably at that time this energy density 
would have contributed to a very fast expansion of the universe. 

These considerations arising from both cosmology and particle physics make it desirable 
to develop a good understanding of the joint evolution of scalar fields and the expansion 
rate of the universe. There is in fact already a large literature on the subject; for a standard 
review see e.g.\ ref.\ \ct{weinberg:2008} and references therein. In the following we explore 
a somewhat different approach to this topic, based on earlier work \ct{jwvh:2002}-\ct{ jwvh:2013b}.
This approach can be used for different purposes; not only to solve the equations of motion, 
but for example also to find the potential and the scale factor necessary to produce a certain 
prescribed evolution of the scalar field. An advantage of this procedure is that in principle it 
makes no use of the slow-roll or any similar approximation. 

\section{Modeling cosmological scalar fields \label{s.2}}

We do not know how many fundamental scalar fields may contribute to the cosmic 
expansion. But we can assume the combined potential of the scalars to form a 
landscape in which there is an effective potential with a well-defined minimum; if there
would be a flat region near the minimum this would give rise to massless or nearly 
massless (pseudo) Goldstone bosons, which we exclude. We also assume that any 
potential symmetries between the scalar fields are broken near the minimum. This 
implies that the effective scalar field fluctuates in a single direction in the scalar 
landscape. This assumption greatly facilitates our analysis. Another simplifying 
assumption, based on well-established observational evidence, is that the universe
is taken to be spatially flat.

The resulting simple model is that of a single homogeneous scalar field $\vf(t)$ in 
a homogeneous and isotropic universe, of which the scale factor $a(t)$ evolves
as determined by the Hubble parameter $H(t) = \dot{a}/a$. The dynamics of these 
parameters is described by the Klein-Gordon equation in a homogeneous and 
isotropic space-time:
\be
\ddot{\vf} + 3H \dot{\vf} + V^{\prime} = 0,
\label{2.1}
\ee
for the scalar field with potential $V[\vf]$, combined with the Einstein equation 
\be
\frac{1}{2}\, \dot{\vf}^2 + V = 3 H^2.
\label{2.2}
\ee
For periods of time in which the scalar field changes monotonically -- e.g., during 
the period of an oscillation between two extreme values -- the evolution of the Hubble 
parameter can be linked to that of the scalar field by considering it as an implicit 
function of time in terms of dependence on the scalar field: $H(t) = H[\vf(t)]$. 
Such a point of view turns out to be remarkably fruitful. First note that 
\be 
\dot{H} = H^{\prime} \dot{\vf},
\label{2.3}
\ee
where the prime denotes differentiation w.r.t.\ the field $\vf$. Differentiation of eq.\ 
(\ref{2.2}) w.r.t.\ time with use of the KG-equation then implies
\be
2 H \dot{H} = - H \dot{\vf}^2. 
\label{2.4}
\ee
Therefore either the universe is static and flat: $H = 0$, or 
\be
\dot{H} = - \frac{1}{2}\, \dot{\vf}^2, \hs{2} \dot{\vf} = - 2 H^{\prime}.
\label{2.5}
\ee
It follows that equation (\ref{2.2}) can be cast in a form determining $H[\vf]$:
\be
3H^2 - 2 H^{\prime\,2} = V.
\label{2.6}
\ee
The points where this description can break down are critical points $\vf_c$ where the 
evolution of the scalar field stops: $\dot{\vf}_c = 0$; at such points $H^{\prime}_c = 0$ and 
$V_c = 3 H_c^2$. These points can be of two kinds: \\
(a) end points of evolution; then 
\be
\ddot{\vf}_c = 4 H_c^{\prime\prime} H^{\prime}_c = 0;
\label{2.7}
\ee
(b) turning points, where 
\be
\ddot{\vf}_c = 4 H_c^{\prime\prime} H^{\prime}_c \neq 0.
\label{2.8}
\ee
As in both cases $H^{\prime}_c = 0$, in case (a) the second derivative of the Hubble parameter 
must be finite: $|H_c^{\prime\prime}| < \infty$; whereas in case (b) this quantity must diverge: 
\be
H_c^{\prime\prime} \propto \frac{1}{H^{\prime}_c} \rightarrow \infty. 
\label{2.9}
\ee
As an example, consider the case of a quadratic potential \ct{jwvh:2013b}
\be
V[\vf] = \ve + \frac{m^2}{2}\, \vf^2.
\label{2.10}
\ee
Then the critical points are $(\vf_c, H_c)$ related by 
\be
3H_c^2 - \frac{m^2}{2}\, \vf_c^2 = \ve.
\label{2.11}
\ee

\bc
\vs{-1}
\scalebox{0.32}{\includegraphics{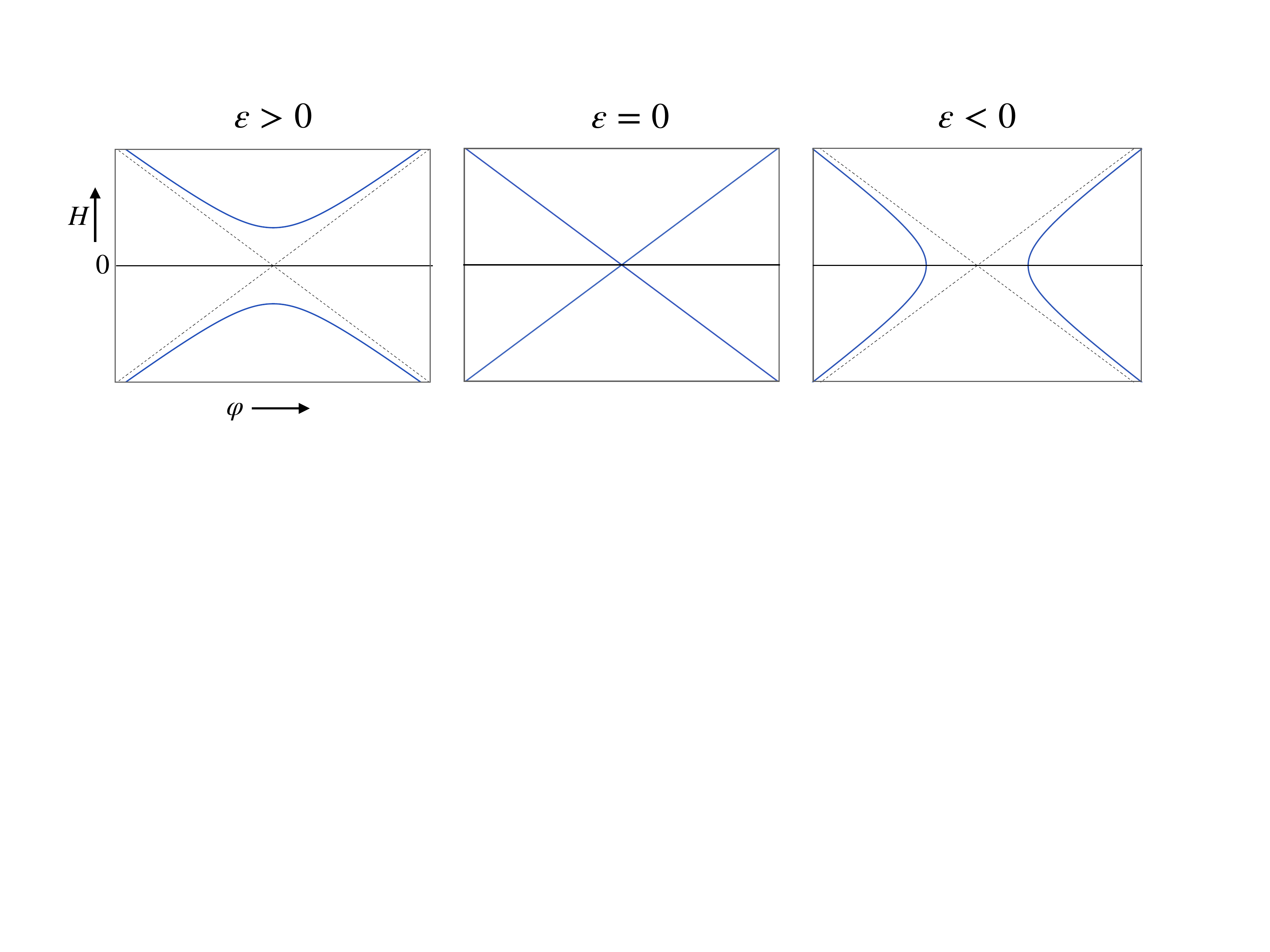}}
\vs{-20}

{\footnotesize Fig.\ 1: Critical points of $H[\vf]$ for scalar field with quadratic potential}
\ec

\nit
Eq.\ (\ref{2.11}) defines a hyperbola, the shape of which is determined by the sign of the vacuum 
energy $\ve$ (fig.\ 1). For $\ve > 0$ and positive $H$, there is a single branch with a 
minimum $H_{min} > 0$ at $\vf = 0$. This minimum represents an endpoint of evolution 
corresponding to a de Sitter universe with a positive cosmological constant; for $\ve = 0$ 
the hyperbola degenerates into the asymptotic straight lines, and the end point of evolution 
becomes Minkowski space with $H = 0$. Finally for $\ve < 0$ the critical curve is a hyperbola 
with two branches, allowing the field to evolve from an expanding regime in which $H > 0$ to 
the contracting regime $H < 0$; the universe will then start to coalesce as soons as $H$ 
crosses into negative values and total collapse becomes inevitable. 

These scenarios are illustrated in fig.\ 2, showing some numerical solutions of equation 
(\ref{2.6}) for the case of the quadratic potential (\ref{2.10}) with $\ve > 0$ (fig.\ 2.a) and 
$\ve < 0$(fig.\ 2.b), respectively.           

\bc
\scalebox{0.2}{\includegraphics{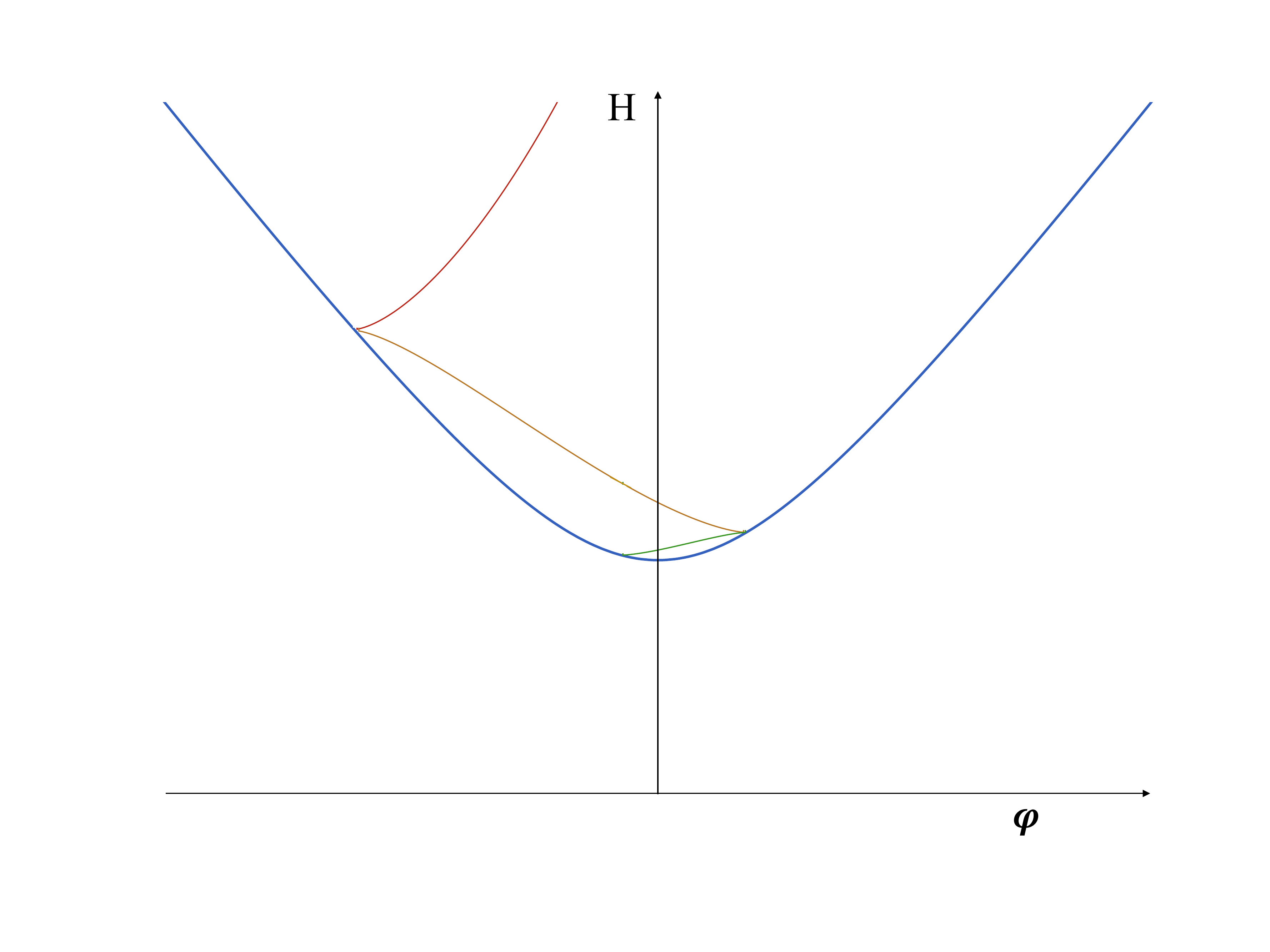}, \includegraphics{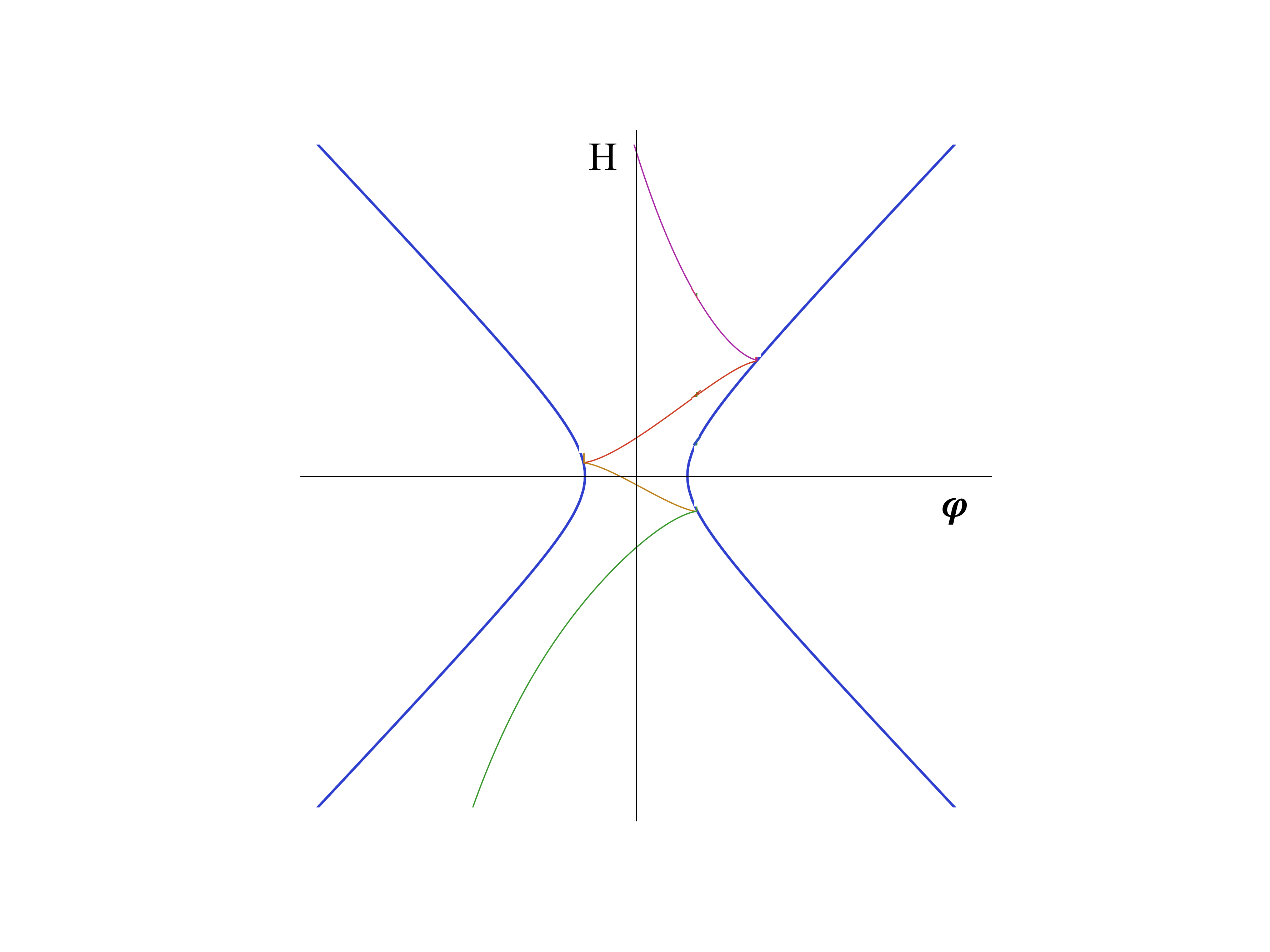}}
\vs{-4}

\[  \hs{1} \mbox{\small a} \hs{17} \mbox{\small b} \]

{\footnotesize Fig. 2: Evolution curves $H[\vf]$ for a quadratic potential with \\
 positive (a) and negative (b) vacuum energy}
\ec

\section{Solving for $H[\vf]$ \label{s.3}}

The specific form of equation (\ref{2.6}) suggests a field redefinition 
\be
u(t) = \sqrt{\frac{3}{2}}\, \vf(t), 
\label{3.1}
\ee
with the result that upon switching the functional dependence of $V$ and $H$ we get
\be
H^2 - H_u^2 = \frac{V}{3}.
\label{3.2}
\ee
For the case of non-negative definite potentials $V[\vf]$ the square root of this equation 
can be found by defining the function $K[u]$ implicitly by
\be
H = \sqrt{\frac{V}{3}}\, \cosh K.
\label{3.3}
\ee
The function $K$ then satisfies the differential equation \ct{jwvh:2013a}
\be
K_u + \frac{V_u}{2V} \cotanh K = \pm 1.
\label{3.4}
\ee
As the left-hand side of this equation is odd in $K$, the two roots are related by 
$K \rightarrow - K$; this results in the same expression for $H$. Therefore we can restrict 
ourselves without loss of generality to the positive sign on the right-hand side. 

A particularly simple example arises for exponential potentials 
\be
V[u] = V_0 e^{2\lb u}, \hs{2} V_0 > 0.
\label{3.5}
\ee
for which
\be
K_u + \lb \cotanh K = 1.
\label{3.6}
\ee
For the special cases $\lb = \pm 1$ the implicit solution $u[K]$ is
\be
4 \lh u - u_0 \rh = 2K - \lb e^{2\lb K}, 
\label{3.10}
\ee
The solutions vanishing at $u = 0$, i.e.\ $K(0) = 0$, have been sketched in fig.\ 3.

\bc
\scalebox{0.23}{\includegraphics{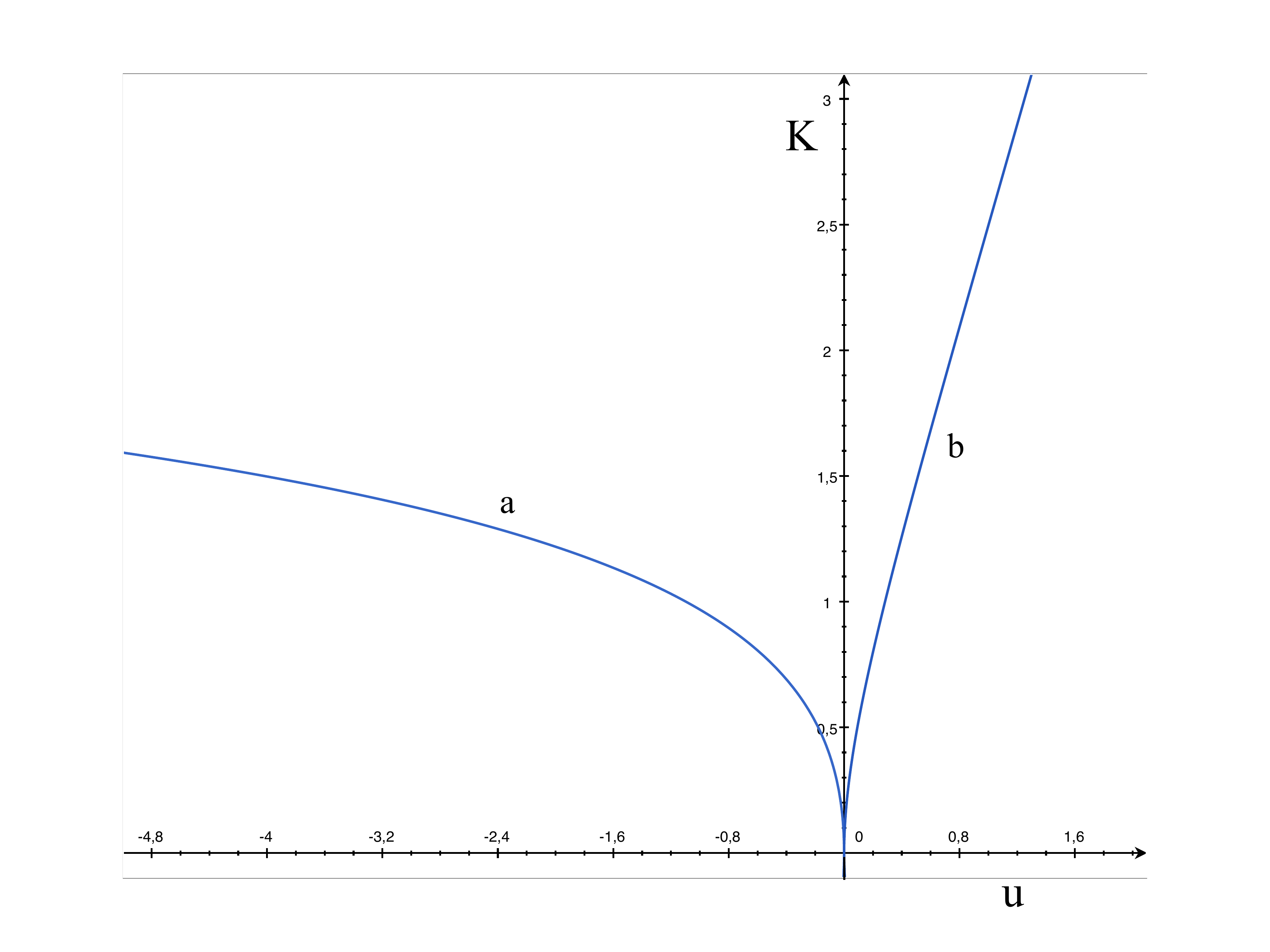}}
\vs{-1}
  
{\footnotesize Fig.\ 3: $K[u]$ for the exponential potential (\ref{3.5}) with $\lb = +1$ (a)
and $\lb = -1$ (b).}
\ec
For general values of $\lb \neq \pm 1$ the implicit solution $u[K]$ of eq.\ (\ref{3.6}) is
\be
\lh1 - \lb^2 \rh \lh u - u_0 \rh = K + \lb \ln \left[ \lh \lb - 1 \rh e^K + \lh \lb + 1 \rh e^{-K} \right].
\label{3.9}
\ee
Furthermore, making use of the identity
\[
H = \frac{\dot{a}}{a} = - \frac{3H_u a_u}{a}, 
\]
it follows from (\ref{3.6}) for all values of $\lb$ that 
\be
K = u + 3\lb \ln \frac{a}{a_0}, 
\label{3.7}
\ee
where $a_0$ is another constant of integration. The definition (\ref{3.3}) then turns into a 
first-order differential equation for the scale factor:
\be
\frac{\dot{a}}{a} = \frac{1}{2} \sqrt{\frac{V_0}{3}} \left[ e^{(\lb + 1) u} \lh \frac{a}{a_0} \rh^{3\lb} 
 - e^{(\lb-1) u} \lh \frac{a}{a_0} \rh^{-3\lb} \right].
\label{3.8}
\ee
 
\section{Reverse engineering of the potential \label{s.4}}

The formalism set down in the previous sections can also be used to solve the reverse 
problem: given a scalar evolution $\vf(t)$, which potential $V[\vf]$ and cosmological 
evolution $H[\vf]$ is it associated with? We demonstrate this for a simple case: exponential
roll-down
\be
\vf(t) = \vf_0 e^{- \og t}.
\label{4.1}
\ee
For large times $t \rightarrow \infty$ the scalar field (\ref{4.1}) tends to
$\vf = \dot{\vf} = \ddot{\vf} = 0$, implying that the only stationary point is an end point. In 
appendix \ref{A.a} we present an example of a field with turning points during its evolution.
For the present case (\ref{4.1}) it follows from the second equation (\ref{2.5}) that 
\be
H[\vf] = h + \frac{\og}{4}\, \vf^2.
\label{4.2}
\ee
The corresponding potential is found directly from eq.\ (\ref{2.6}):
\be
V[\vf] = \ve + \frac{m^2}{2}\, \vf^2 + \frac{\lb}{4}\, \vf^4,
\label{4.3}
\ee
with
\be
\ve = 3h^2, \hs{1} m^2 = 3 \og h - \og^2, \hs{1} \lb = \frac{3\og^2}{4}.
\label{4.4}
\ee
The potential is reflection symmetric: $V[\vf] = V[-\vf]$, but if $3 \og h - \og^2  = - \mu^2 < 0$ 
(when $h$ and $\ve$ are small) this symmetry is broken in the minima of the potential at 
$\vf_{\pm} = \pm \sqrt{\mu^2/\lb}$; see fig.\ 4. 

\bc
\scalebox{0.22}{\includegraphics{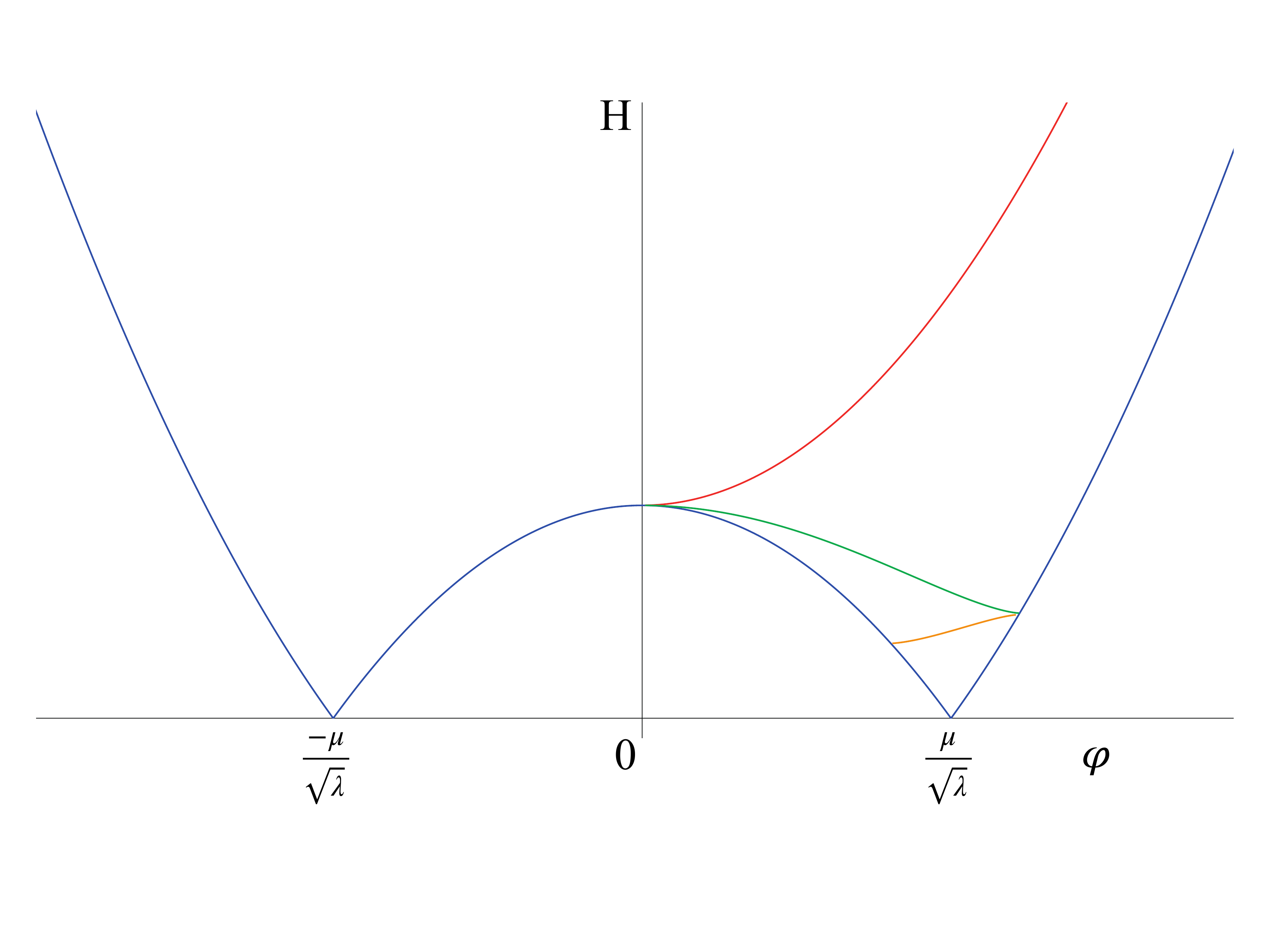}}
\vs{-2}

{\footnotesize \begin{tabular}{rl} Fig.\ 4: & Exponential roll-down for broken quartic potential. \\ 
 & The local maximum at $\vf = 0$ is an unstable end point, after \\
 & which the roll-down continues to one  of the minima at $V = 0$. \end{tabular}}
\ec

\nit
In that case the solution (\ref{4.1}) rolls down to the unstable local maximum of the potential 
at $\vf = 0$. From this maximum the scalar field can continue to evolve in either direction 
to the minimum of the potential. The minimal potential energy density is reached for 
\be
\vf_{min}^2 = \frac{4}{3} \lh 1 - \frac{3h}{\og} \rh \hs{1} \Rightarrow \hs{1} 
V_{min} =- \frac{\og}{3} \lh \og - 6h \rh.
\label{4.5}
\ee
Excluding $\og \leq 0$, the potential is non-negative definite only in the range $0 < \og \leq 6h$.
For the minimal potential energy density to equal the present cosmological energy density 
$V_{min} = 3 H^2_{today} \sim 10^{-120}$ (in Planck units), it requires to great accuracy
\be
\og = 6h, \hs{2} \vf^2_{\min} = \frac{2}{3}.
\label{4.5.a}
\ee 
The critical lines $H[\vf_c]$ and the particular scalar-field trajectory are displayed in fig.\ 4. 

Given the scalar field (\ref{4.1}) and the Hubble parameter (\ref{4.2}) the evolution of
the scale factor during this exponential decay phase can straighforwardly be computed to be 
\ct{jwvh:2002}
\be
a(t) = a(0) e^{ ht + \frac{1}{8} \vf_0^2 \lh 1 - e^{-2\og t} \rh}.
\label{4.6}
\ee
An interesting feature of this solution is, that in addition to eventual exponential expansion
with Hubble constant $h$, it also describes an early phase of super-exponential expansion 
during a time $\tau = 1/2\og$ which leads to an additional number of $e$-folds determined 
by the initial scalar amplitude:
\be
\Del N = \frac{1}{8}\, \vf_0^2.
\label{4.7}
\ee
It follows that the the initial value of the Hubble parameter is 
\be
H_0 = h \lh 1 + 12 \Del N \rh.
\label{4.8}
\ee
This corresponds to the Planck energy if
\be
3 H_0^2 = 3 h^2 \lh 1 + 12 \Del N \rh^2 = 1 \hs{1} \Rightarrow \hs{1} 
h = \frac{1}{\sqrt{3}\, (1 + 12 \Del N)}.
\label{4.9}
\ee
In the standard scenario the number of $e$-folds during inflation is required to be 
$\Del N \geq 60$; in Planck times\footnote{We define $\tau_P= \sqrt{8\pi G \hbar/c^5}
= 0.27 \times 10^{-42}$ s.} $\tau_P$ this would imply
\be
h \leq \frac{1}{830\, \tau_P}, \hs{1} \mbox{or} \hs{1} \og \leq \frac{1}{140\, \tau_P}.
\label{4.10}
\ee
This determines the time in which the scalar field rolls down to the unstable maximum 
at $\vf = 0$. How long it can stay there before rolling down to the minimal value 
(\ref{4.5.a}) depends on the characteristics of the scalar fluctuations. 

The scalar fluctuations $\chi = \vf - \vf_{min}$ near the minimum of the potential are 
described by the potential 
\be
V[\chi] = h^2 \lh 18 \chi^2 + 18 \sqrt{\frac{3}{2}}\, \chi^3 + \frac{27}{4}\, \chi^4 \rh,
\label{4.11}
\ee
which implies that for $\Del N = 60$ the mass of these Higgs particles is
\be
m_{\chi}^2 = 36 h^2 = \frac{1}{12 (\Del N)^2} \sim 2 \times 10^{-5}.
\label{4.12}
\ee
In standard units the Higgs mass $m_{\chi}$ then is of the order of the GUT scale 
$10^{16}$ GeV/$c^2$. Note that we obtain this number without any input of standard 
model particle physics.

\section{Discussion \label{s.5}}

In this overview of isotropic and homogeneous cosmic scalar field dynamics we have
shown how to solve the combined Einstein-Klein-Gordon field equations, by using the 
field itself as a time variable. Considerable qualitative information is obtained from such 
a formulation, and in a number of interesting cases they can be solved exactly without 
making use of approximations. 

In the standard analysis of cosmic scalar field dynamics use is often made of the 
slow-roll approximation, valid when
\be
\eps = - \frac{\dot{H}}{H^2} < 1. 
\label{5.1}
\ee
In our treatment, taking account of eqs.\ (\ref{2.5}) and (\ref{2.6}), this regime 
corresponds to 
\be
V = 3H^2 - 2H^{\prime\,2} , \hs{1} \mbox{and} \hs{1}  
\eps = \frac{2H^{\prime\,2}}{H^2} < 1, 
\label{5.2}
\ee
which leads to the condition
\be
2H^2 < V < 3H^2.
\label{5.3}
\ee

\bc
\scalebox{0.22}{\includegraphics{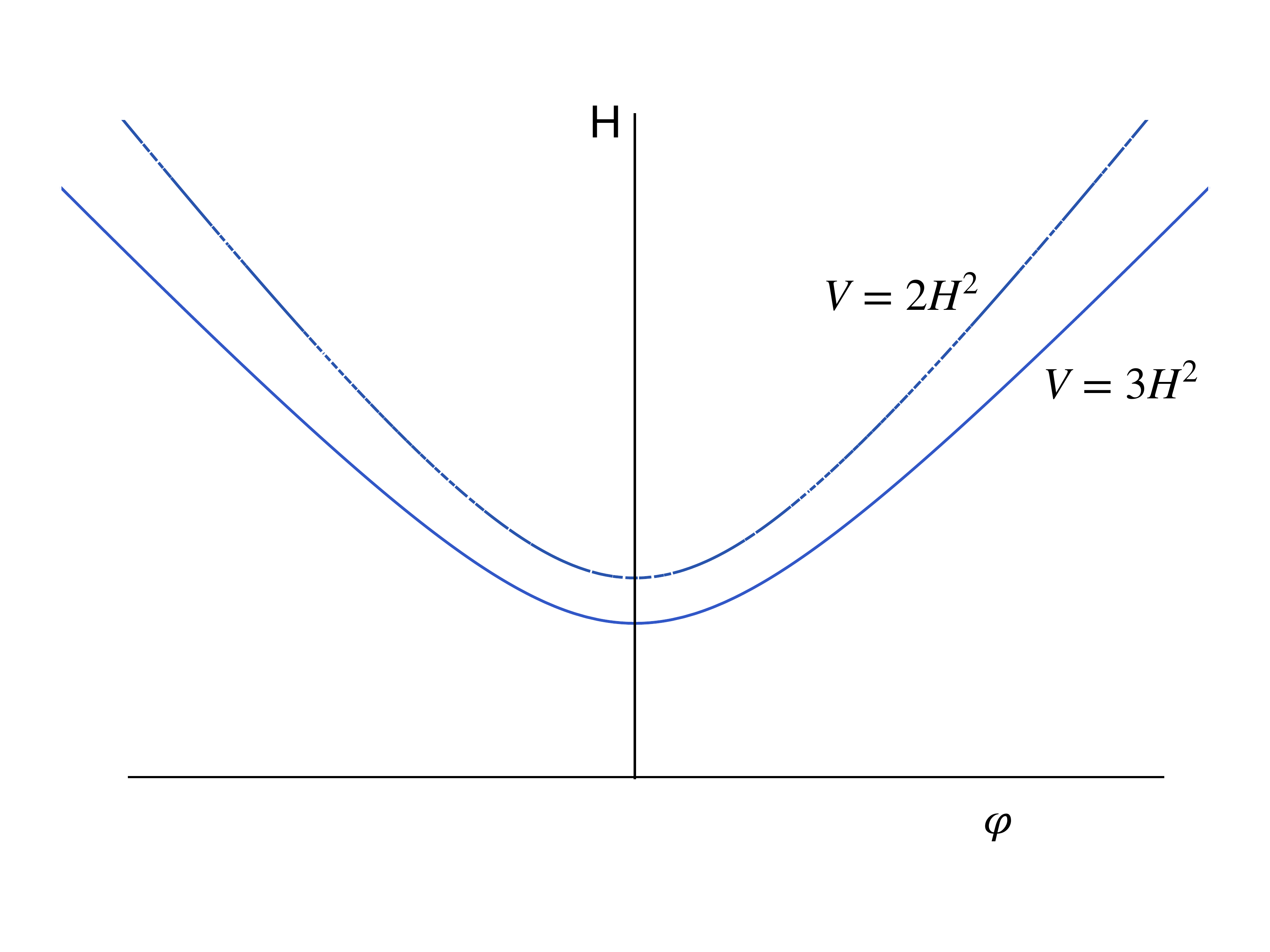}}

{\footnotesize Fig.\ 5: The slow-roll regime for the quadratic potential (\ref{2.10}). }
\ec

\nit
This regime is sketched for the quadratic potential with $\ve > 0$ in fig.\ 5. The 
domain where the condition (\ref{5.3}) holds is seen to be that where the field is 
close to a stationary point, i.e.\ where $\dot{\vf}$ becomes small such that its value 
stays almost constant. During such times the kinetic energy is small and the potential 
behaves more or less like a cosmological constant, whereas during periods in which 
the kinetic terms are large $H$ diminishes faster and little is contributed to the 
expansion of the universe. The procedure described in this paper does not make use 
of the slow-roll restriction to this regime, and as shown in sections \ref{s.2} and \ref{s.3} 
solutions of the Einstein-Klein-Gordon equations can be obtained valid in the full range 
where these classical equations hold. However, in specific applications where it is 
useful to work in the slow-roll regime our central equation (\ref{2.6}) can be solved by 
a perturbative method \ct{roest:2014} ; this is explained in appendix \ref{A.b}.

We have also seen that Higgs-type potentials can give rise to a limited period of 
super-exponential expansion of the early universe.  Remarkably, if this period is 
equated to a period of inflation by 60 $e$-folds, as favored by cosmological 
data, the mass of the associated Higgs particle lies naturally in the range of
Grand Unification. 

Finally we remark that beyond the single-scalar scenario the evolution of a universe
with several dynamical homogeneous and istotropic scalar fields has been considered 
e.g.\ in refs.\ \ct{nibbelink:2002} and \ct{kerner:2014} and references therein.

\appendix

\section{The oscillating scalar field \label{A.a}}

In section \ref{s.4} we discussed the construction of a scalar potential associated 
with exponential decay of the scalar field. The end point of this decay is by 
necessity a static field. In the present section we present a different example in
which the field has an indefinite number of stationary points, all of which are 
turning points. This is the case for an oscillating scalar field \ct{jwvh:2013b}
\be
\vf(t) = \vf_0 \cos \og t,
\label{a.1}
\ee 
which has turning points at all times $\og t_n = n \pi$. Such a field evolution requires 
\be
H' = - \frac{1}{2}\, \dot{\vf} = \frac{\og}{2} \sqrt{\vf_0^2 - \vf^2}.
\label{a.2}
\ee
As a result 
\be
H^{\prime\prime} = - \frac{1}{2}\, \frac{\og \vf}{\sqrt{\vf_0^2 - \vf^2}},
\label{a.3}
\ee
which diverges at all turning points, as expected.
By integration of (\ref{a.2}) it is straightforward to establish that 
\be
\ba{lll}
H[\vf] & = & \dsp{ H_0 - \frac{1}{4}\, \og \vf_0^2 \arccos \lh \frac{\vf}{\vf_0} \rh + \frac{1}{4}\, \og \vf
 \sqrt{\vf_0^2 - \vf^2} }\\
 & & \\
 & = & \dsp{ H_0 - \frac{1}{4}\, \og^2 \vf_0^2 t + \frac{1}{8}\, \og \vf_0^2 \sin 2 \og t. }
\ea
\label{a.4}
\ee
The signs here are such, that although $H(t)$ oscillates, on average it decreases. 
The corresponding solution for the scale factor is 
\be
a(t) = a(0)  e^{H_0 t - \frac{1}{8}\, \og^2 \vf_0^2 t^2 - \frac{\vf_0^2}{16} \cos 2 \og t}.
\label{a.5}
\ee
Thus $a(t)$ displays gaussian behaviour, slightly modified by an oscillating amplitude; see fig.\ 6. 

\bc
\vs{-2}
\scalebox{0.22}{\includegraphics{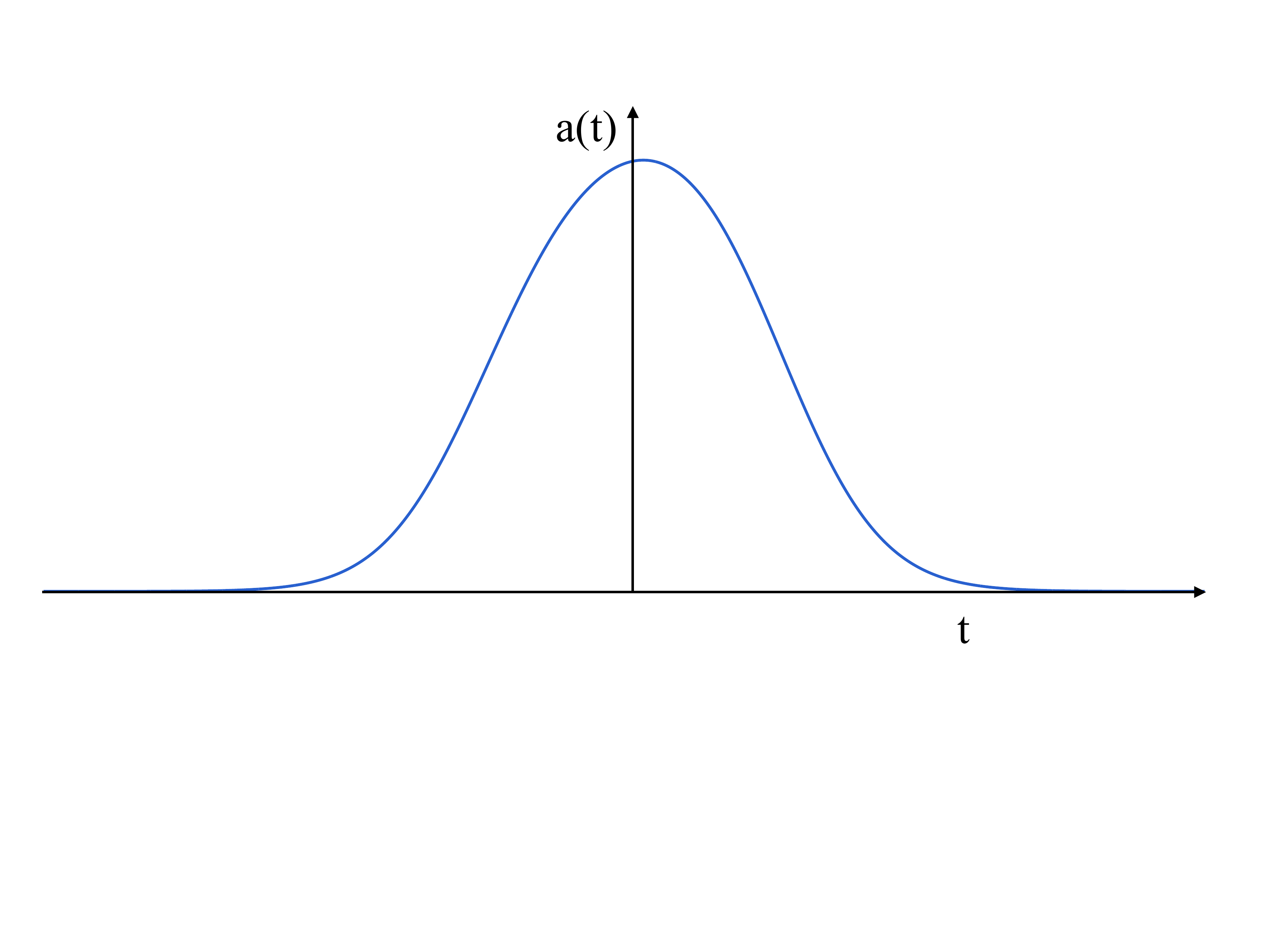}}
\vs{-9}

{\footnotesize Fig.\ 6: Scalefactor driven by oscillating scalar field}
\ec

\nit
This implies that a universe dominated by such a scalar field does not expand indefinitely,
but reaches a maximal expansion at time
\be
t = \frac{4H_0^2}{\og^2 \vf_0^2},
\label{a.6}
\ee
and then collapses again. The scalar potential which gives rise to such a solution of the 
Einstein-Klein-Gordon equations is 
\be
V[\vf] = 3 \lh H_0 - \frac{1}{4}\, \og \vf_0^2 \arccos \lh \frac{\vf}{\vf_0} \rh + \frac{1}{4}\, \og \vf
 \sqrt{\vf_0^2 - \vf^2} \rh^2 - \frac{\og^2}{2} \lh \vf_0^2 - \vf^2 \rh.
\label{a.7}
\ee
We conclude that undamped oscillations of a strong scalar field can not be accomodated in a 
realistic cosmological model.  

\section{Pertubative solutions in the slow-roll regime \label{A.b}}

Given the scalar potential $V[\vf]$ one can try to construct solutions for $H[\vf]$ in the 
neighborhood of a critical point $\vf_c$, where the slow-roll conditions hold, by series 
expansion. Indeed, as discussed in sect.\ \ref{s.5} near critical points $H$ is a slowly 
changing function of $\vf$ and these regions, where the kinetic scalar terms are small, 
provide the major contributions to the expansion of the universe. In this appendix we 
develop this perturbative approach; we illustrate the general procedure with the 
quadratic potential of sect.\ \ref{s.2}. 

At a critical point $\vf_c$ the Hubble parameter is stationary: $H'[\vf_c] = 0$. For the 
construction of a solution in a neighborhood of this critical point we may try to expand 
$H'[\vf]$ in terms of $\hvf = \vf - \vf_c$. We then have to distinguish between the two 
cases disussed in sect.\ \ref{s.2}. \\
(a) If the critical point is an end point then $H^{\prime\prime}[\vf_c]$ is finite and $H'[\vf]$ 
can be expanded into a convergent power series of the form 
\be
\ld H'\right|_{\vf = \vf_c + \hvf} = \sum_{n = 1}^{\infty} \frac{1}{n!}\, \eta_n\, \hvf^n 
 = \eta_1\, \hvf + \frac{1}{2}\, \eta_2\, \hvf^2 +  ...
\label{b.1}
\ee
which vanishes at $\hvf = 0$, with 
\be
H^{\prime\prime}[\vf] = \sum_{n = 0}^{\infty} \frac{1}{n!}\, \eta_{n+1}\, \hvf^n = \eta_1 + \eta_2\, \hvf 
 + \frac{1}{2}\, \eta_3\, \hvf^2 + ...
\label{b.2}
\ee 
convergent near $\vf_c$ such that $H'_c H^{\prime\prime}_c = 0$. By integration of (\ref{b.1})
\be
H[\vf] = \eta_0 + \sum_{n = 1}^{\infty} \frac{1}{(n+1)!}\, \eta_n\, \hvf^{n+1} 
 = \eta_0 + \frac{1}{2}\, \eta_1\, \hvf^2 + ...,
\label{b.3}
\ee
with $\eta_0$ a constant of integration. The coefficients in the expansion (\ref{b.3}) can 
now be fixed by applying eq.\ (\ref{2.6}). 

We can illustrate this by the example of the quadratic potential (\ref{2.10}). For this potential 
with $\ve > 0$ the end point of the scalar evolution is $\vf_c = 0$ and $\hvf = \vf$. Then 
the above expansions imply
\be 
3 H^2 - 2 H^{\prime\,2} = 3 \eta_0^2 + \lh 3 \eta_0 \eta_1 - 2 \eta_1^2 \rh \vf^2 + 
 \lh \eta_0 \eta_2 - 2 \eta_1 \eta_2 \rh \vf^3 + ... = \ve + \frac{m^2}{2}\, \vf^2.
\label{b.4}
\ee
By comparing the coefficients on both sides 
\be
3 \eta_0^2 = \ve, \hs{1} 3 \eta_0 \eta_1 - 2 \eta_1^2 = m^2, \hs{1} \eta_2 \lh \eta_0 - 2 \eta_1 \rh = 0, 
 \hs{1} ...
\label{b.5}
\ee
This hierarchy of equations determines all coefficients $\eta_n$ sequentially. Note 
however, that a solution for $\eta_1$ exists for positive $m^2$ only when 
\be
 9 \eta_0^2 = 3 \ve \geq 8 m^2. 
\label{b.6}
\ee
It turns out that if this condition is not satisfied, there is no solution which reaches
the stationary end point in a finite number of oscillations; as we will see below the 
scalar field keeps oscillating around the minimum of the potential with ever decreasing 
amplitude. 

As we have already discussed, in the case $\ve < 0$ the universe collapses, there 
is no finite end point of evolution and the procedure above does not apply. Indeed, 
in the neighborhood of $H = 0$ there exists solutions of the type 
\be
H[\vf] = h_1 \hvf + \frac{1}{2}\, h_2 \hvf^2 + ..., \hs{2} 
H'[\vf] = h_1 + h_2 \hvf + \frac{1}{2}\, h_3 \hvf^2 + ...
\label{3.7}
\ee
with $H[\vf_c] = 0$ and
\be
2 h_1^2 = - \ve, \hs{1} 4h_1 h_2 = - m^2 \vf_c, \hs{1} 
6 h_1^2 - 4 h_2^2 - 4 h_1h_3 = m^2, ...
\label{b.8}
\ee
Thus it crosses into the regime of a collapsing universe with $H < 0$ at $\vf = \vf_c$. \\
(b) The other type of solutions to be discussed concerns those near a turning point 
$\vf_c$ with divergent $H_c^{\prime\prime}$.  In this case eq.\ (\ref{2.8}) implies that
$H'_c H^{\prime\prime}_c = \kg$, a non-zero finite constant. Therefore near $\vf_c$ 
\be
H^{\prime\,2}[\vf] = 2 \kg \hvf + ... 
\label{b.9}
\ee
As a result instead of an expansion in powers of $\hvf$ we get an expansion in powers of 
$|\hvf|^{1/2}$. For convenience of discussion, we assume that $\hvf > 0$, i.e.\ the field 
approaches the critical value $\vf_c$ from the right. We can then make the expansion 
\be
H'[\vf] = \sum_{n=1}^{\infty} \kg_n\, \hvf^{n/2} = \kg_1 \hvf^{1/2} + \kg_2\, \hvf + \kg_3\, \hvf^{3/2} 
 + \kg_4 \hvf^2 + ...
\label{b.10}
\ee
It follows that 
\be
H[\vf] = \kg_0 + \sum_{n = 1}^{\infty} \frac{2\kg_n}{n+2}\, \hvf^{(n+2)/2} = 
 \kg_0 + \frac{2\kg_1}{3}\, \hvf^{3/2} + \frac{\kg_2}{2}\, \hvf^2 + ...,
\label{b.11}
\ee
and as a result
\be
3 H^2 - 2 H^{\prime\,2} = 3 \kg_0^2 - 2 \kg_1^2 \hvf + 4 \lh \kg_0 \kg_1 
 - \kg_1 \kg_2 \rh \hvf^{3/2} + \lh 3 \kg_0 \kg_2 - 2 \kg_2^2 - 4 \kg_1 \kg_3 \rh \hvf^2 + ..., 
\label{b.12}
\ee
which is to be compared with the potential $V[\vf_c + \hvf]$ in order to establish the 
solution for the coefficients $\kg_n$. 

For the example of the quadratic potential (\ref{2.10}) with $\ve > 0$ we get the explicit results 
\be
\ba{l}
\dsp{ 3 \kg_0^2 = \ve + \frac{m^2}{2}\, \vf_c^2, \hs{1} - 2 \kg_1^2 = m^2 \vf_c, \hs{1} 
\kg_1 \lh \kg_0 - \kg_2 \rh = 0, }\\
 \\
\dsp{ 6 \kg_0 \kg_2 - 4 \kg_2^2 - 8 \kg_1 \kg_3 = m^2, \hs{1} 
3 \kg_0 \kg_3 - 5 \kg_2 \kg_3 - 5 \kg_1 \kg_4 = 0, \hs{1} ... }
\ea
\label{b.13}
\ee
Note that this requires $\vf_c < 0$, which is consistent with our assumption that  the field
approaches the turning point from the right. Also note, that as the turning point is not at the 
minimum of the potential: $\vf_c \neq 0$, it follows that $\kg_1 \neq 0$ and $\kg_2 = \kg_0$. 
The equations on the second line then become 
\be
\kg_3 = \frac{2 \kg_0^2 - m^2}{8 \kg_1}, \hs{1} \kg_4 = - \frac{2 \kg_0 \kg_3}{5\kg_1}, \hs{1} ...
\label{b.14}
\ee
These equations have solutions for any $\ve \geq 0$, and therefore apply also in the 
regime disallowed by (\ref{b.6}). However, in that regime the limit $\vf_c \rightarrow 0$ 
can not be reached, and solutions where $\vf_c$ becomes an end point cannot be 
constructed; the scalar fields keeps oscillating with ever decreasing amplitude. Indeed 
as $\vf$ passes through $\vf_c$ coming from the right, $\dot{\vf}$ changes from 
negative to positive with finite $\ddot{\vf}_c = \kg_1^2/2 > 0$. Because of the relation 
(\ref{2.5}) the Hubble parameter then changes slope as a function of $\vf$ from positive 
to negative; therefore with $\vf$ increasing after passing the turning point, $H[\vf]$ must 
decrease. In terms of the solution (\ref{b.11}) for $H[\vf]$ this means, that the solution 
with $\kg_1 >  \sqrt{m^2 |\vf_c|/2} > 0$ turns into the solution with 
$\kg_1= - \sqrt{m^2 |\vf_c| /2} < 0$. Then similarly the coefficient $\kg_3$ and all 
higher coefficients $\kg_{2n+1}$ change sign, which amounts to replacing 
$\hvf^{(2n+1)/2} \rightarrow - \hvf^{(2n+1)/2}$ in equations (\ref{b.10}) and (\ref{b.11}).

Once the series expansions (\ref{b.10}) and (\ref{b.11}) have been obtained, one also 
knows $\dot{\vf} = - 2H'$, and it is possible to compute the number of $e$-folds of expansion
in the regime between $H_1$ and $H_2$ by 
\be
N = \int_1^2 H dt = - \frac{1}{2}\, \int_1^2 \frac{H}{H'}\, d \vf.
\label{b.15}
\ee

\vs{3}

\end{document}